\documentclass[journal,12pt,onecolumn,draftclsnofoot]{IEEEtran}

\usepackage{amsmath}
\usepackage{graphicx,psfrag} 
\usepackage{amsmath,amsthm,amsfonts,amssymb,bm} 
\usepackage[ruled,vlined]{algorithm2e}
\usepackage{algpseudocode}
\usepackage{xcolor}
\usepackage{bbm}             
\usepackage{tabularx}        
\usepackage{framed}
\usepackage{subfigure}
\usepackage[latin1]{inputenc}
\allowdisplaybreaks[3]

\newcommand{\bs}{\mathbf}

\begin{document}

\title{
Expectation-Maximization-Aided Hybrid Generalized Expectation Consistent for Sparse Signal Reconstruction
}

\author{
    Qiuyun~Zou,
    Haochuan~Zhang$^*$,
    and~Hongwen~Yang

\thanks{
Q. Zou and H. Yang are with Beijing University of Posts and Telecommunications,
Beijing 100876, China (email: qiuyunzou@qq.com; yanghong@bupt.edu.cn).
H. Zhang is with Guangdong University of Technology, Guangzhou 510006, China (email: haochuan.zhang@gdut.edu.cn).
 $^*$Corresponding author: H. Zhang.
}
}

\maketitle

\begin{abstract}
The reconstruction of sparse signal is an active area of research. Different from a typical i.i.d. assumption, this paper considers a non-independent prior of group structure. For this more practical setup, we propose EM-aided HyGEC, a new algorithm to address the stability issue and the hyper-parameter issue facing the other algorithms. The instability problem results from the ill condition of  the transform matrix, while the unavailability of the hyper-parameters is a ground truth that their values are not known beforehand. The proposed algorithm is built on the paradigm of HyGAMP (proposed by Rangan et al.) but we replace its inner engine, the GAMP, by a matrix-insensitive alternative, the GEC, so that the first issue is solved. For the second issue, we take expectation-maximization as an outer loop, and together with the inner engine HyGEC, we learn the value of the hyper-parameters. Effectiveness of the proposed algorithm is also verified by means of numerical simulations.
\end{abstract}


\section{Introduction}

Recently the high-dimensional signal recovery of structured sparse signal of the generalized linear model involving a linear mixing space and a componentwise mapping channel has a wide range of applications in many engineering fields such as compressive sensing \cite{schniter2010turbo,rangan2017hybrid}, image processing \cite{som2012compressive}, and wireless communication \cite{liu2018massive,zou2020message,zou2020low}, etc. To estimate the signal of interest, several principles were developed.  Among them, the greedy pursuit algorithms such as matching pursuit (MP) and orthogonal matching pursuit (OMP) \cite{pati1993orthogonal} can be regarded as a variant of least square (LS), in which the residual error of each iteration was projected onto an atom or a sub-hyperplane. A method based on the maximum likelihood principle can be found in \cite{tipping2001sparse}, where the sparse Bayesian learning algorithm was proposed by assuming Gaussian prior. Nevertheless, both of them didn't utilize the true prior information.

As a solution to this inference problem, the Bayesian estimator can fully use the prior information. To implement approximate Bayesian inference iteratively, there exists two kinds of algorithms, i.e., approximate message passing (AMP) \cite{donoho2010message} and expectation propagation (EP) \cite{minka2001family}. AMP and its extensions approximate the loopy belief propagation (LBP) \cite{kschischang2001factor} based on bipartite graph by performing Gaussian approximation and Taylor expansion. The high-dimensional problem represented by factor graph can be decomposed into a set of smaller problems by delivering the messages between factor nodes and variable nodes via their edges. On the other hand, the EP algorithm derived from assumed density filter (ADF) was used to approximate factorable distribution by minimizing Kullback-Leibler (KL) divergence, also named relative entropy. The EP algorithm is very close to the vector AMP (VAMP) \cite{rangan2019vector}, expectation consistent (EC) \cite{opper2005expectation}, and orthogonal AMP (OAMP) \cite{ma2017orthogonal}. In \cite{rangan2017hybrid}, Rangan et al. proposed hybrid generaized AMP (HyGAMP) by splitting the factor graph into two part, in which the standard message passing was performed in strong edges and the GAMP was run in weak edges (linear mixing).
However, HyGAMP for structured sparse signal has the following limitations. Firstly, HyGAMP fails to converge when the measurement matrix is non-zero mean and ill-condition. Secondly, they need to know exactly what value each (hyper-)parameter is (say, the sparse rate $\rho$). In practice, such information is difficult to obtain. For instance, in massive connectivity \cite{zou2020message, zou2020low}, each sparse rate is determined by an individual user, rather than by a base station or any centralized controller. In addition, the existing works \cite{metzler2018expectation,fletcher2017learning,fletcher2017rigorous,fletcher2019plug} pertaining to VAMP with hyper-parameter estimation were not for group sparse signal.

To address these issues, also to investigate more possibilities of the paradigm proposed by Rangan et al. \cite{rangan2017hybrid}, we in this paper consider the enhancement in two aspects:
1) replacing the inner engine by a more adaptable generalized expectation consistent (GEC) technique \cite{fletcher2016expectation,he2017generalized}, so that the stability issue could be avoided;
2) embedding the resultant estimator into a larger framework of expectation-maximization (EM), where the true sparse rate could be learned iteratively.

\section{Problem Formulation}
Consider the signal recovery problem below, where $\bs{x}\in \mathbb{R}^{N}$ is the signal to recover, $\bs{y}\in \mathbb{R}^{M}$ is the observation, and $\bs{H}\in \mathbb{R}^{M\times N}$ is the measurement matrix that linearly transforms $\bs{x}$ into $\bs{z}=\bs{Hx}$. This $\bs{z}$ is mapped randomly into $\bs{y}$ according to the transition distribution of 
\begin{align}
\mathcal{P}(\bs{y}|\bs{z})=\prod_{m=1}^M\mathcal{P}(y_m|z_m), \ \text{s.t.},\ \bs{z}=\bs{Hx}.
\label{Equ:System}
\end{align}
Different from the classical setup of i.i.d. prior, we allow here the prior distribution of  $\bs{x}$ to have some  \emph{structured dependency}, i.e., dividing the elements of $\bs{x}$ into $K$ non-overlapping groups, $\bs{x}=\{\bs{x}_k\}_{k=1}^K$, where $\sum_{k=1}^K N_{k}=N$ and $N_k$ is the number of elements in the $k$-th group, we  allow the $N_{k}$ elements of the $k$-th group to be arbitrarily dependent within that group, but retain their independence of any elements from any other groups.
In \cite{rangan2017hybrid}, such a structured dependency was exemplified via group sparsity, and this paper follows that convention.
In the group sparsity context, the activity of random elements $\{x_{kj}\}$ within the $k$-th group is controlled by a single binary indicator $\xi_k\in \{0,1\}$, and that is carried out by: $\forall j  \in \{1,\ldots, N_k\}$
\begin{align*}
\mathcal{P}(x_{kj}|\xi_k)
&=\xi_k\mathcal{P}_{\textsf{X}}(x_{kj})+(1-\xi_k)\delta(x_{kj}),
\end{align*}
where $\delta(\cdot)$ is the Dirac delta function. The indicator $\xi_k$ itself is a Bernoulli r.v. following
$\mathcal{P}\{\xi_k=1\}=1-\mathcal{P}\{\xi_k=0\}=\rho_k$, with $\mathcal{P}\{\cdot\}$ here denoting the probability of a random event, and $\rho_k$ being the sparse rate.
Since different groups are (assumed to be) independent, one could choose a different sparse rate for each group, but for the sake of simplicity, we use w.l.o.g. an uniform setup: $\rho_1=\rho_2=\cdots=\rho$. In all, it reads
\begin{align*}
\mathcal{P}(\bs{x}|\bs{\xi})
&=
\prod_{k=1}^K \prod_{j=1}^{N_k} \mathcal{P}(x_{kj}|\xi_k),
\ \text{s.t.},\ \xi_k \sim \text{Bern}(\rho).
\end{align*}


We are interested in obtaining an approximate to the output of a classical MMSE estimator for each $x_{kj}$ \cite{kay1993fundamentals}:
\begin{align}
\hat{x}_{kj}=\int x_{kj}\mathcal{P}(x_{kj}|\bs{y})\text{d}x_{kj},
\end{align}
where $\mathcal{P}(x_{kj}|\bs{y})$ is a marginal posterior distribution of the joint posterior $\mathcal{P}(\bs{x},\boldsymbol{\xi}|\bs{y})$
\begin{align}
\mathcal{P}(x_{kj}|\bs{y})&=\int_{\bs{x}_{\backslash kj}}\int_{\boldsymbol{\xi}}\mathcal{P}(\bs{x},\boldsymbol{\xi}|\bs{y})\text{d}\boldsymbol{\xi}\text{d}\bs{x}_{\backslash kj}
,
\\
\mathcal{P}(\bs{x},\boldsymbol{\xi}|\bs{y})
&\propto \mathcal{P}(\bs{y}|\bs{x},\boldsymbol{\xi})\mathcal{P}(\bs{x}|\boldsymbol{\xi})\mathcal{P}(\boldsymbol{\xi})
,
\label{Equ:Joint}
\\
& \propto
\int_{\bs{z}} \mathcal{P}(\bs{y}|\bs{z}) \delta(\bs{z}-\bs{Hx})\text{d}\bs{z}
\prod_{k=1}^K\prod_{j=1}^{N_k} \left[\mathcal{P}(x_{kj}|\xi_k)\mathcal{P}(\xi_k)\right]
,\nonumber
\end{align}
where $\bs{x}_{\backslash kj}$ is equal to $\bs{x}$ except its element $x_{kj}$.
The reason we are interested in an approximate solution is that obtaining an exact $\hat{x}_{kj}$ is difficult, because of a high complexity in the exact computation of the marginals, as $N,M \to \infty$.

\section{The Proposed Algorithm: EM-Aided HyGEC}
This section implements the algorithm, referring to the EM-aided HyGEC following a convention of its precedents.  The new algorithm is given in Algorithm \ref{alg:HyGEC+EM}, and before proceeding to its deduction, we remind the readers of its two technical steps. One is that currently there are two types of nodes and messages in the network, scalar and vector; the other is the output of the HyGEC is a point estimate but what the EM require as its input is a distribution function.
The handling of these technical steps constitutes the following subsections.
\begin{figure}
\centering
\includegraphics[width=0.5\textwidth]{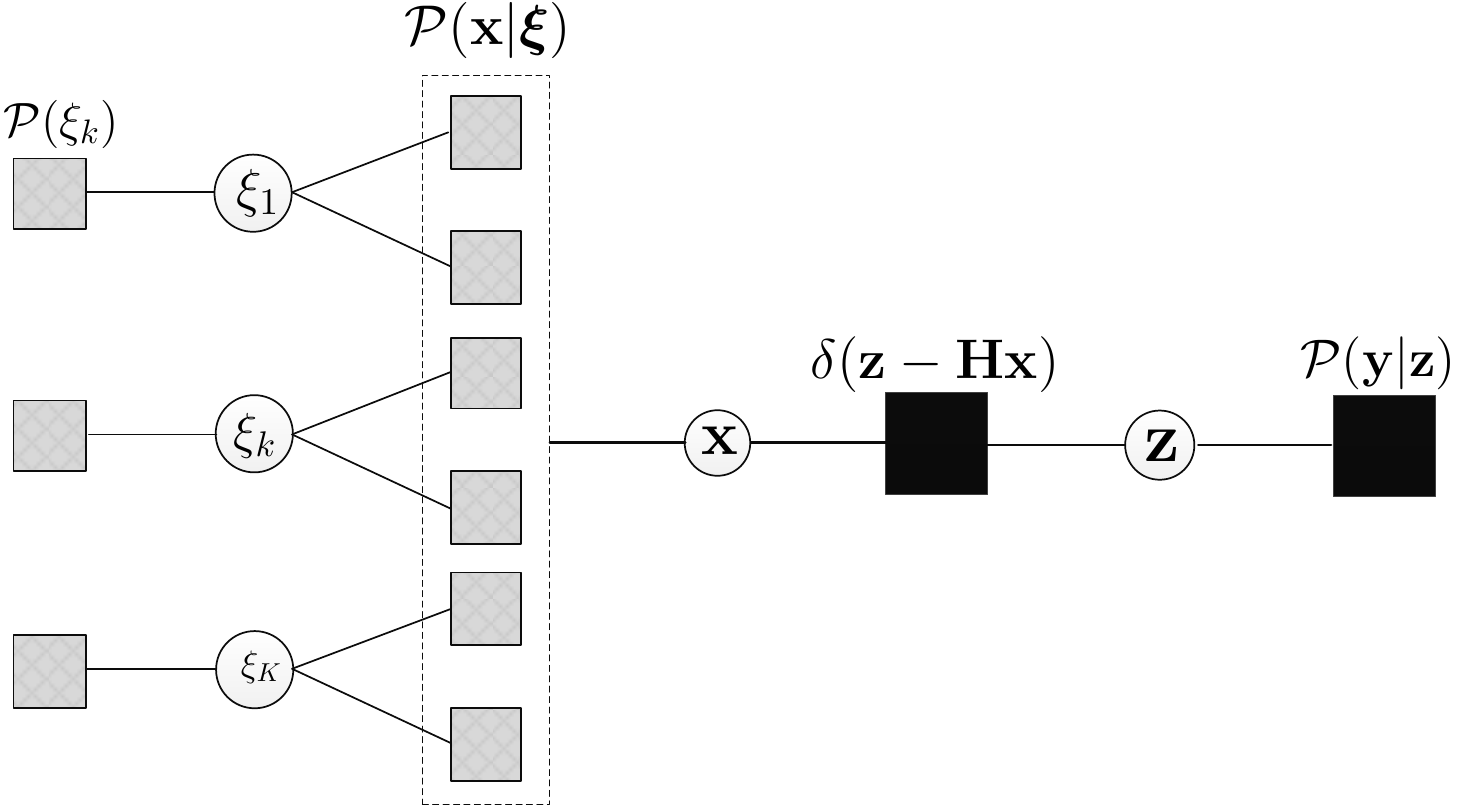}
\caption{A ``hybrid'' factor graph relating to the (conditional) joint PDF of (\ref{Equ:Joint}): each gray-shaded node on the l.h.s. represents a scale-value function or variable, while each black-highlighted node on the r.h.s. represents a vector-value function or variable.
}
\label{Fig1:FG}
\end{figure}

\begin{algorithm}[!t]
	\caption{EM-Aided HyGEC}
	\label{alg:HyGEC+EM}
	{
		\small
		\begingroup
		\addtolength{\jot}{-0em}
		\textbf{0. Input}: $\bs{H}$, $\bs{Y}$\\
		\textbf{1. Initialize}: $\rho(0) \in (0,1)$ \\
		\textbf{2. Iterate} (outer loop, EM): \\
		\For{$t=0,\cdots,T$}
		{			
			E-step (inner engine, HyGEC): \\  \quad $(\bs{m}_{\text{x}}^{\text{lik}}(t),\bs{v}_{\text{x}}^{\text{lik}}(t),\hat{\boldsymbol{\rho}}(t),\hat{\bs{x}}^{\text{pos}}(t))=\text{HyGEC}(\rho(t))$;\\
	        M-step (parameter update): \\ \quad
	        compute $\rho(t+1)$ according to (\ref{Equ:rho}) and (\ref{Equ:Pik});\\
			\textbf{until} $\|\hat{\bs{x}}^{\text{pos}}(t+1)-\hat{\bs{x}}^{\text{pos}}(t)\|<\epsilon$ or $t>T$;
		}
		\endgroup
		\textbf{3. Output}:  $\hat{\bs{x}}^{\text{pos}}$ and $\rho$
	}
\end{algorithm}

\subsection{Inner Engine: HyGEC Using a Postulated Sparse Rate}
In Fig.~\ref{Fig1:FG}, we depict the factor graph relating to the joint distribution $\mathcal{P}(\bs{x},\boldsymbol{\xi}|\bs{y})$ of (\ref{Equ:Joint}).
This factor graph is not a standard defined one: all its nodes on the left hand side are in a scalar form, while all nodes on the right hand side are in the vector form.
Therefore, messages circulating between ``$\mathcal{P}(\boldsymbol{\xi})\leftrightarrow\boldsymbol{\xi}\leftrightarrow\mathcal{P}(\bs{x}|\boldsymbol{\xi})$'' are scalar, and those between  ``$p(\bs{x}|\boldsymbol{\xi})\leftrightarrow\delta(\bs{z}-\bs{Hx})\leftrightarrow\bs{z}\leftrightarrow\mathcal{P}(\bs{y}|\bs{z})$'' are vector.
For the scalar (left) messages, we adopt the standard sum-product LBP rules \cite{kschischang2001factor} in their updating; for the vector (right) ones, we use the GEC manner which was also termed as the EP-like rules in \cite{zou2018concise,meng2015expectation}.
The message updating rule for the middle variable node $\bs{x}$ needs some special treatment, in order to serve our purpose of exchanging information between the two parts. Details on the derivation of HyGEC is presented in Appendix \ref{Sec1}. We present only its pseudocode in Algorithm \ref{alg:HyGEC}.
It is also worthy of noting that we also assume w.l.o.g. the prior $\mathcal{P}_{\textsf{X}}(x_{kj})$ takes a Gaussian form and thus the conditional distribution is
$
\mathcal{P}(x_{kj}|\xi_k)=\xi_k\mathcal{N}(x_{kj}|0,\sigma^2_X)+(1-\xi_k)\delta(x_{kj}).
$

\begin{algorithm}[!t]
	\caption{
		  $(\bs{m}_{\text{x}}^{\text{lik}},\bs{v}_{\text{x}}^{\text{lik}},\hat{\boldsymbol{\rho}},\hat{\bs{x}}^{\text{pos}})=\text{HyGEC}(\rho)$
	}
	\label{alg:HyGEC}
	{
		\small
		\begingroup
		\addtolength{\jot}{-0.4em}
		\textbf{0. Input}: $\rho$\\
        \textbf{1. Definition}:
        \begin{align*}
        \mathcal{P}(z|m,v)&=\frac{\mathcal{P}(y|z)\mathcal{N}(z|m,v)}{\int \mathcal{P}(y|z)\mathcal{N}(z|m,v) \text{d}z}\\
        \mathcal{P}(x|m,v;\hat{\rho})&=\frac{\mathcal{P}(x;\hat{\rho})\mathcal{N}(x|m,v)}{\int \mathcal{P}(x;\hat{\rho})\mathcal{N}(x|m,v)\text{d}x}\\
        \mathcal{P}(x;\hat{\rho})&=\hat{\rho}\mathcal{N}(x|0,\sigma_X^2)+(1-\hat{\rho})\delta(x)
        \end{align*}
		\textbf{2. Init}: $\bs{m}_{\text{z}}^{\text{pri}}=\bs{0}$, $\bs{v}_{\text{z}}^{\text{pri}}=P_z\bs{1}$, $\bs{m}_{\text{x}}^{\text{pri}}=\bs{0}$,  $\bs{v}_{\text{x}}^{\text{pri}}=\rho\bs{1}$\\
		\textbf{3. Iterate}: \\
		\For{$t=1,\cdots,T$}
		{
			{\setlength\abovedisplayskip{1pt}
				\setlength\belowdisplayskip{1pt}
				\begin{align*}
				\hat{\bs{z}}^{\text{pos}}&=\mathbb{E}\left\{\bs{z}|\bs{m}_{\text{z}}^{\text{pri}},\bs{v}_{\text{z}}^{\text{pri}}\right\}
				\\
				\bs{v}_{\text{z}}^{\text{pos}}&=\text{Var}\left\{\bs{z}|\bs{m}_{\text{z}}^{\text{pri}},\bs{v}_{\text{z}}^{\text{pri}}\right\}
				\\
				\bs{v}_{\text{z}}^{\text{lik}}&=\bs{1}\oslash (\bs{1}\oslash \bs{v}_{\text{z}}^{\text{pos}}-\bs{1}\oslash \bs{v}_{\text{z}}^{\text{pri}})
				\\
				\bs{m}_{\text{z}}^{\text{lik}}&=\bs{v}_{\text{z}}^{\text{lik}}\odot (\hat{\bs{z}}^{\text{pos}}\oslash \bs{v}_{\text{z}}^{\text{pos}}-\bs{m}_{\text{z}}^{\text{pri}}\oslash \bs{v}_{\text{z}}^{\text{pri}})
				\\
				\bs{Q}_{\text{x}}&=\left(\bs{H}^{\text{T}}\text{Diag}(\bs{1}\oslash \bs{v}^{\text{lik}}_{\text{z}})\bs{H}+\text{Diag}(\bs{1}\oslash \bs{v}_{\text{x}}^{\text{pri}})\right)^{-1}
				\\
				\hat{\bs{x}}^{\text{pos2}}&=\bs{Q}_{\text{x}}\left(\bs{H}^{\text{T}}\text{Diag}(\bs{1}\oslash \bs{v}^{\text{lik}}_{\text{z}})\bs{m}^{\text{lik}}_{\text{z}}+\bs{m}_{\text{x}}^{\text{pri}}\oslash \bs{v}_{\text{x}}^{\text{pri}}\right)
				\\
				\bs{v}_{\text{x}}^{\text{pos2}}&=\text{diag}(\bs{Q}_{\text{x}})
				\\
				\bs{v}_{\text{x}}^{\text{lik}}&=\bs{1}\oslash (\bs{1}\oslash \bs{v}_{\text{x}}^{\text{pos2}}-\bs{1}\oslash \bs{v}_{\text{x}}^{\text{pri}})
				\\
				\bs{m}_{\text{x}}^{\text{lik}}&=\bs{v}_{\text{x}}^{\text{lik}}\odot (\hat{\bs{x}}^{\text{pos2}}\oslash \bs{v}_{\text{x}}^{\text{pos2}}-\bs{m}_{\text{x}}^{\text{pri}}\oslash \bs{v}_{\text{x}}^{\text{pri}})
				\\
				\hat{\bs{x}}^{\text{pos}}&=\mathbb{E}\left\{\bs{x}|\bs{m}_{\text{x}}^{\text{lik}},\bs{v}_{\text{x}}^{\text{lik}};\hat{\boldsymbol{\rho}}\right\}
				\\
				\bs{v}_{\text{x}}^{\text{pos}}&=\text{Var}\left\{\bs{x}|\bs{m}_{\text{x}}^{\text{lik}},\bs{v}_{\text{x}}^{\text{lik}};\hat{\boldsymbol{\rho}}\right\}
				\\
				\bs{v}_{\text{x}}^{\text{pri}}&=\bs{1}\oslash (\bs{1}\oslash \bs{v}_{\text{x}}^{\text{pos}}-\bs{1}\oslash  \bs{v}_{\text{x}}^{\text{lik}})
				\\
				\bs{m}_{\text{x}}^{\text{pri}}&=\bs{v}_{\text{x}}^{\text{pri}}\odot (\hat{\bs{x}}^{\text{pos}}\oslash \bs{v}_{\text{x}}^{\text{pos}}-\bs{m}_{\text{x}}^{\text{lik}}\oslash  \bs{v}_{\text{x}}^{\text{lik}})
				\\
				\bs{Q}_{\text{x}}&=\left(\bs{H}^{\text{T}}\text{Diag}(\bs{1}\oslash \bs{v}^{\text{lik}}_{\text{z}})\bs{H}+\text{Diag}(\bs{1}\oslash \bs{v}_{\text{x}}^{\text{pri}})\right)^{-1}
				\\
				\hat{\bs{x}}^{\text{pos}2}&=\bs{Q}_{\text{x}}\left(\bs{H}^{\text{T}}\text{Diag}(\bs{1}\oslash \bs{v}^{\text{lik}}_{\text{z}})\bs{m}^{\text{lik}}_{\text{z}}+\bs{m}_{\text{x}}^{\text{pri}}\oslash \bs{v}_{\text{x}}^{\text{pri}}\right)
				\\
				\hat{\bs{z}}^{\text{pos1}}&=\bs{H}\hat{\bs{x}}^{\text{pos}2}
				\\
				\bs{v}^{\text{pos1}}_{\text{z}}&=\text{diag}(\bs{H}\bs{Q}_{\text{x}}\bs{H}^{\text{T}})\\
				\bs{v}_{\text{z}}^{\text{pri}}&=\bs{1}\oslash (\bs{1}\oslash \bs{v}^{\text{pos1}}_{\text{z}}-\bs{1}\oslash \bs{v}_{\text{z}}^{\text{lik}})\\
				\bs{m}_{\text{z}}^{\text{pri}}&=\bs{v}_{\text{z}}^{\text{pri}}\odot (\hat{\bs{z}}^{\text{pos1}}\oslash \bs{v}^{\text{pos1}}_{\text{z}}-\bs{m}_{\text{z}}^{\text{lik}}\oslash \bs{v}_{\text{z}}^{\text{lik}})\\
				\text{LLR}_{k\leftarrow j}^{\xi}
				&=\log \frac{\mathcal{N}(0|m_{\text{x},kj}^{\text{lik}},\sigma_X^2+v_{\text{x},kj}^{\text{lik}})}{\mathcal{N}(0|m_{\text{x},kj}^{\text{lik}},v_{\text{x},kj}^{\text{lik}})}\\
				\text{LLR}_{k}^{\xi}&=\log \frac{\rho}{1-\rho}+\sum_{i=1}^{N_k}\text{LLR}_{k\leftarrow i}^{\xi}\\
				\text{LLR}_{k\rightarrow j}^{\xi}&=\text{LLR}_{k}^{\xi}-\text{LLR}_{k\leftarrow j}^{\xi}\\
				\hat{\rho}_{kj}&=1-\frac{1}{1+\exp (\text{LLR}_{k\rightarrow j}^{\xi})}
				\end{align*}
				\textbf{until} $\|\hat{\bs{x}}^{\text{pos}}(t+1)-\hat{\bs{x}}^{\text{pos}}(t)\|^2< \epsilon$ or $t>T$.
		}}
		\endgroup
		\textbf{3. Output}: $(\bs{m}_{\text{x}}^{\text{lik}},\bs{v}_{\text{x}}^{\text{lik}},\hat{\boldsymbol{\rho}},\hat{\bs{x}}^{\text{pos}})$
	}
	
\end{algorithm}

\subsection{Outer Iteration: EM to Learn the Actual Sparse Rate}
In last subsection, the HyGEC requires a known sparse rate $\rho$; however, this parameter in practice is not known yet, and here we rely on EM to learn its true value.
EM \cite{vila2013expectation} is an iterative technique that increases a lower bound on the likelihood function $\mathcal{P}(\bs{y};\rho)$  for the hyper-parameter estimate at each iteration. For an arbitrary distribution $q(\bs{x})$, we have
\begin{align}
\ln \mathcal{P}(\bs{y};\rho)
&=\int q(\bs{x})\ln \mathcal{P}(\bs{y};\rho)\text{d}\bs{x}\\
&=\int q(\bs{x})\ln \left[\frac{\mathcal{P}(\bs{x},\bs{y};\rho)}{q(\bs{x})}\frac{q(\bs{x})}{\mathcal{P}(\bs{x}|\bs{y};\rho)}\right]\text{d}\bs{x}\\
&=\underbrace{\mathbb{E}_{q}\left\{\ln \frac{\mathcal{P}(\bs{x},\bs{y};\rho)}{q(\bs{x})}\right\}}_{\overset{\triangle}{=}\ \text{ELBO}(q,\bs{y};\rho)} +\underbrace{\mathcal{D}_{\text{KL}} \left[q||\mathcal{P}\right]}_{\geq 0},
\end{align}
where $\mathbb{E}_{q}\{\cdot\}$ denotes the expectation over $q(\bs{x})$, $\mathcal{D}_{\text{KL}}$ refers to the Kullback-leibler (KL) divergence between $q(\bs{x})$ and posterior $\mathcal{P}(\bs{x}|\bs{y};\rho)$, and ``ELBO'' is the abbr of evidence lower bound. Since the KL divergence is non-negative, it implies that $\text{ELBO}(q,\bs{y};\rho)$ is the low bound of $\log \mathcal{P}(\bs{y};\rho)$. The EM algorithm can be divided into two steps: E-step) finding a $q(\bs{x})$ to minimize the KL divergence $\mathcal{D}(q(\bs{x})\|\mathcal{P}(\bs{x}|\bs{y};\rho))$ given the parameter $\rho=\rho(t)$; M-step) finding the parameter $\rho$ to maximize the $\text{ELBO}(q,\bs{y};\rho)$ based on $q(\bs{x},t)$.

In E-step of EM algorithm, given the parameter $\rho(t)$, we aim at finding the $q(\bs{x})$ to minimize the KL divergence $\mathcal{D}_{\text{KL}}(q(\bs{x},t)\|\mathcal{P}(\bs{x}|\bs{y};\rho(t)))$. To this end, we apply the HyGEC algorithm as shown in Algorithm \ref{alg:HyGEC} to approximate the posterior $\mathcal{P}(\bs{x}|\bs{y};\rho(t))$.
As we mentioned before,  the output of HyGEC is point estimate, but what we need here is a density function $q(\bs{x},t)$. A bridge closing this gap is to use the point estimates as parameters of that density, i.e.,
\begin{align}
q(\bs{x},t)=\frac{\mathcal{P}(\bs{x};\hat{\boldsymbol{\rho}}(t))\mathcal{N}(\bs{x}|\bs{m}_{\text{x}}^{\text{lik}}(t),\bs{v}_{\text{x}}^{\text{lik}}(t))}{\int \mathcal{P}(\bs{x}; \hat{\boldsymbol{\rho}}(t))\mathcal{N}(\bs{x}|\bs{m}_{\text{x}}^{\text{lik}}(t),\bs{v}_{\text{x}}^{\text{lik}}(t)) \text{d}\bs{x}},
\end{align}
where $\mathcal{P}(\bs{x};\hat{\boldsymbol{\rho}}(t))=\{\hat{\mathcal{P}}(x_{kj};\hat{\rho}_{kj}(t)),\forall k,j\}$. Note that the approximate posterior $q(\bs{x},t)$ are independent element-wisely.

In M-step of EM algorithm, given an approximated posterior $q(\bs{x},t)$, we update the prior parameter $\rho(t)$ by maximizing $\text{ELBO}(q,\bs{y};\rho)$, i.e.,
\begin{align}
\rho(t+1)&=\underset{\rho}{\arg \max}\ \text{ELBO}(q,\bs{y};\rho).
\label{Equ:M-step}
\end{align}
The $\text{ELBO}(q,\bs{y};\rho)$ can further be expanded as
\begin{align}
\text{ELBO}(q,\bs{y};\rho)&=\mathbb{E}_q\{\ln \mathcal{P}(\bs{y}|\bs{x})+\ln \mathcal{P}(\bs{x};\rho) \}+\mathbb{E}_q\{\ln q(\bs{x})\},
\end{align}
where $\mathcal{P}(\bs{x};\rho)$ is expressed as
\begin{align}
\mathcal{P}(\bs{x};\rho)
&=\prod_{k=1}^{K}\int \prod_{j=1}^{N_k}\mathcal{P}(x_{kj}|\xi_k)\mathcal{P}(\xi_k)\text{d}\xi_k\\
&=\prod_{k=1}^{K}\left[\rho\prod_{j=1}^{N_k}\mathcal{P}_{\textsf{X}}(x_{kj})+(1-\rho)\prod_{j=1}^{N_k}\delta(x_{kj})\right].
\end{align}
Note that only the term $\mathbb{E}_q\{\ln \mathcal{P}(\bs{y}|\bs{x})\}$ in $\text{ELBO}(q,\bs{y};\rho)$ is related to $\rho$. Then (\ref{Equ:M-step}) can be further written as
\begin{align}
\rho(t+1)
&=\underset{\rho}{\arg \max} \int q(\bs{x},t)\ln\mathcal{P}(\bs{x};\rho)\text{d}\bs{x}\\
&=\underset{\rho}{\arg \max} \ \sum_{k=1}^K \int q(\bs{x}_k,t)\ln \mathcal{P}(\bs{x}_k,\rho)\text{d}\bs{x}_k.
\label{Equ:rho1}
\end{align}
We calculate the partial derivation of $\ln\mathcal{P}(\bs{x}_k;\rho)$ w.r.t. $\rho$
\begin{align}
\!\!\!\!\!\!\!\! \frac{\partial \ln\mathcal{P}(\bs{x}_k;\rho)}{\partial \rho}&=\frac{\prod_{j=1}^{N_k}\mathcal{P}_{\textsf{X}}(x_{kj})-\prod_{j=1}^{N_k}\delta(x_{kj})}{\rho\prod_{j=1}^{N_k}\mathcal{P}_{\textsf{X}}(x_{kj})+(1-\rho)\prod_{j=1}^{N_k}\delta(x_{kj})}\\
&=
\begin{cases}
\frac{1}{\rho}   &\bs{x}_k\ne \bs{0}\\
-\frac{1}{1-\rho}  &\bs{x}_k=\bs{0}
\end{cases}.
\end{align}
We define an Euclidean ball in $\mathbb{R}^{N_k}$ having the form $\mathcal{B}=\{\bs{z}|\|\bs{z}\|\leq \epsilon\}$ and it complement $\overline{\mathcal{B}}=\mathbb{R}^{N_k}\backslash \mathcal{B}$. From (\ref{Equ:rho1}), in the limit $\epsilon\rightarrow 0$, we have
\begin{align}
\!\!\! \frac{1}{\rho}\sum_{k=1}^K\int_{\bs{x}_k\in \overline{\mathcal{B}}}q(\bs{x},t)\text{d}\bs{x}_k=\frac{1}{1-\rho}\sum_{k=1}^K\int_{\bs{x}_k\in \mathcal{B}}q(\bs{x}_k,t)\text{d}\bs{x}_k,
\label{Equ:rho2}
\end{align}
in which
\begin{align}
\int_{\bs{x}_k\in \overline{\mathcal{B}}}q(\bs{x}_k,t)\text{d}\bs{x}_k
& =\prod_{j=1}^{N_k}\frac{1}{1+\frac{1-\hat{\rho}_{kj}(t)}{\hat{\rho}_{kj}(t)}
\frac{\mathcal{N}(0|m_{\text{x},kj}^{\text{lik}}(t),v_{\text{x},kj}^{\text{lik}}(t))}{\mathcal{N}(0|m_{\text{x},kj}^{\text{lik}}(t),v_{\text{x},kj}^{\text{lik}}(t)+\sigma_X^2)}}\\
&\overset{\triangle}{=}\pi_k(\bs{m}_{\text{x},k}^{\text{lik}}(t),\bs{v}_{\text{x},k}^{\text{lik}}(t),\boldsymbol{\hat{\rho}}_{k}(t)).
\label{Equ:Pik}
\end{align}
where $\bs{m}_{\text{x},k}^{\text{lik}}(t)=\{{m}_{\text{x},kj}^{\text{lik}}(t),\forall j\}$ and $\bs{v}_{\text{x},k}^{\text{lik}}(t)=\{{v}_{\text{x},kj}^{\text{lik}}(t),\forall j\}$.

From (\ref{Equ:rho2}), we have
\begin{align}
\rho(t+1)&=\frac{1}{K}\sum_{k=1}^{K}\pi_k(\bs{m}_{\text{x},k}^{\text{lik}}(t),\bs{v}_{\text{x},k}^{\text{lik}}(t),\boldsymbol{\hat{\rho}}_{k}(t)).
\label{Equ:rho}
\end{align}
Totally, the EM-aided HyGEC algorithm for structured signal recovery with unknown sparse rate is summarized in Algorithm~\ref{alg:HyGEC+EM}.

\section{Simulation and Discussion}
In this section, the simulations are presented to validate the performance of the proposed algorithm compared to the existing competing algorithm. We particularize the generalized linear model (\ref{Equ:System}) as the following quantization model
\begin{align}
\bs{y}=\textsf{Q}(\bs{Hx}+\bs{w}),
\end{align}
where $\bs{w}$ is additive white Gaussian noise (AWGN) with variance $\sigma_w^2$ and $\textsf{Q}(\cdot)$ denotes an analog to digital converter (ADC). In particular, as the number of quantization bits becomes sufficiently large, the quantization model reduces to the standard linear model. The closed form of ($\hat{\bs{z}}^{\text{pos}}, \bs{v}_{\text{z}}^{\text{pos}}$) can be obtained following a similar procedure as given by \cite[Appendix A]{wen2015bayes}.

\begin{figure}
\centering
\includegraphics[width=0.6\textwidth]{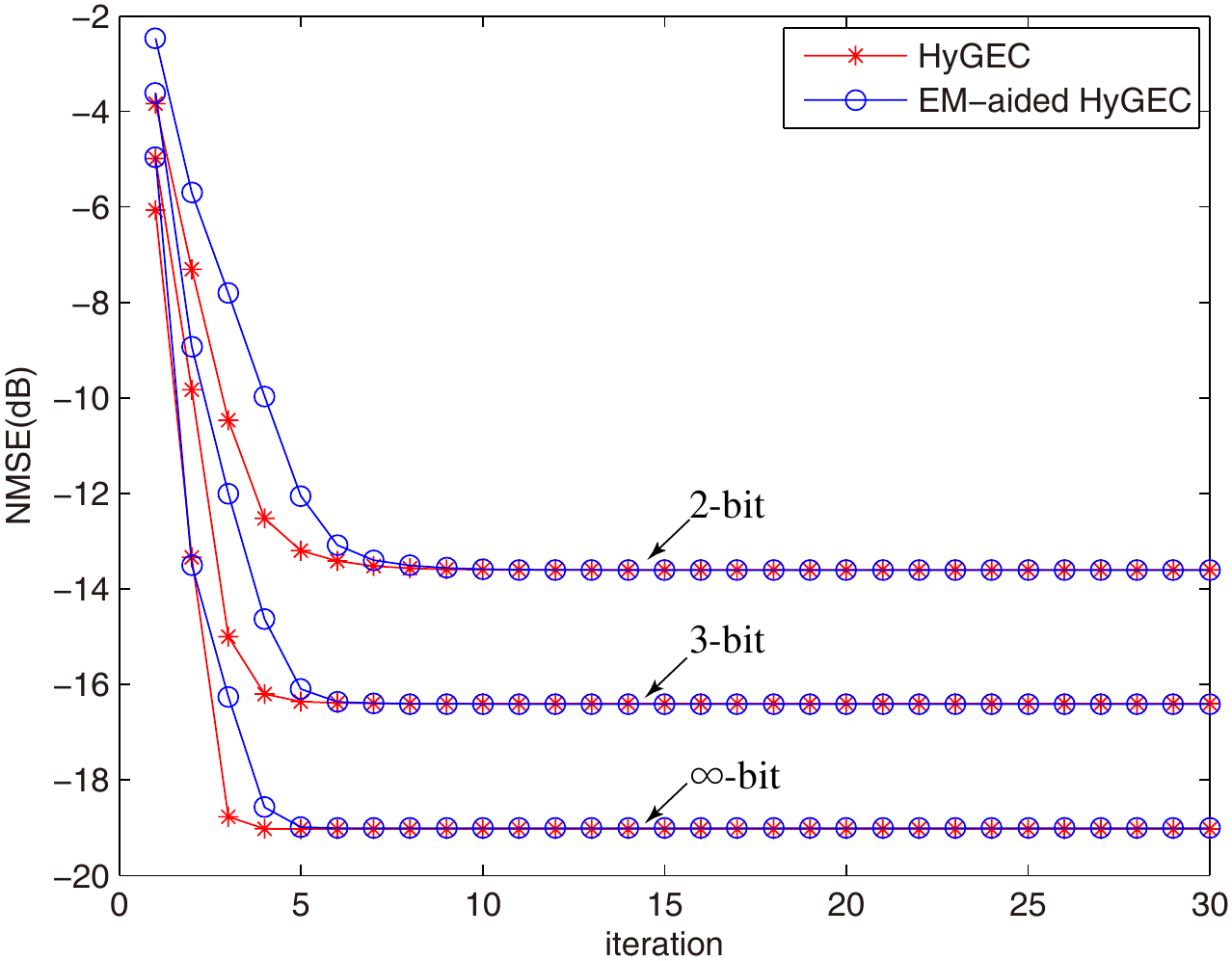}
\caption{ Per-iteration behavior of HyGEC (known $\rho$) and EM-aided HyGEC (unknown $\rho$).
}
\label{Fig:Sim1}
\vspace{+0.3cm}
\centering
\includegraphics[width=0.6\textwidth]{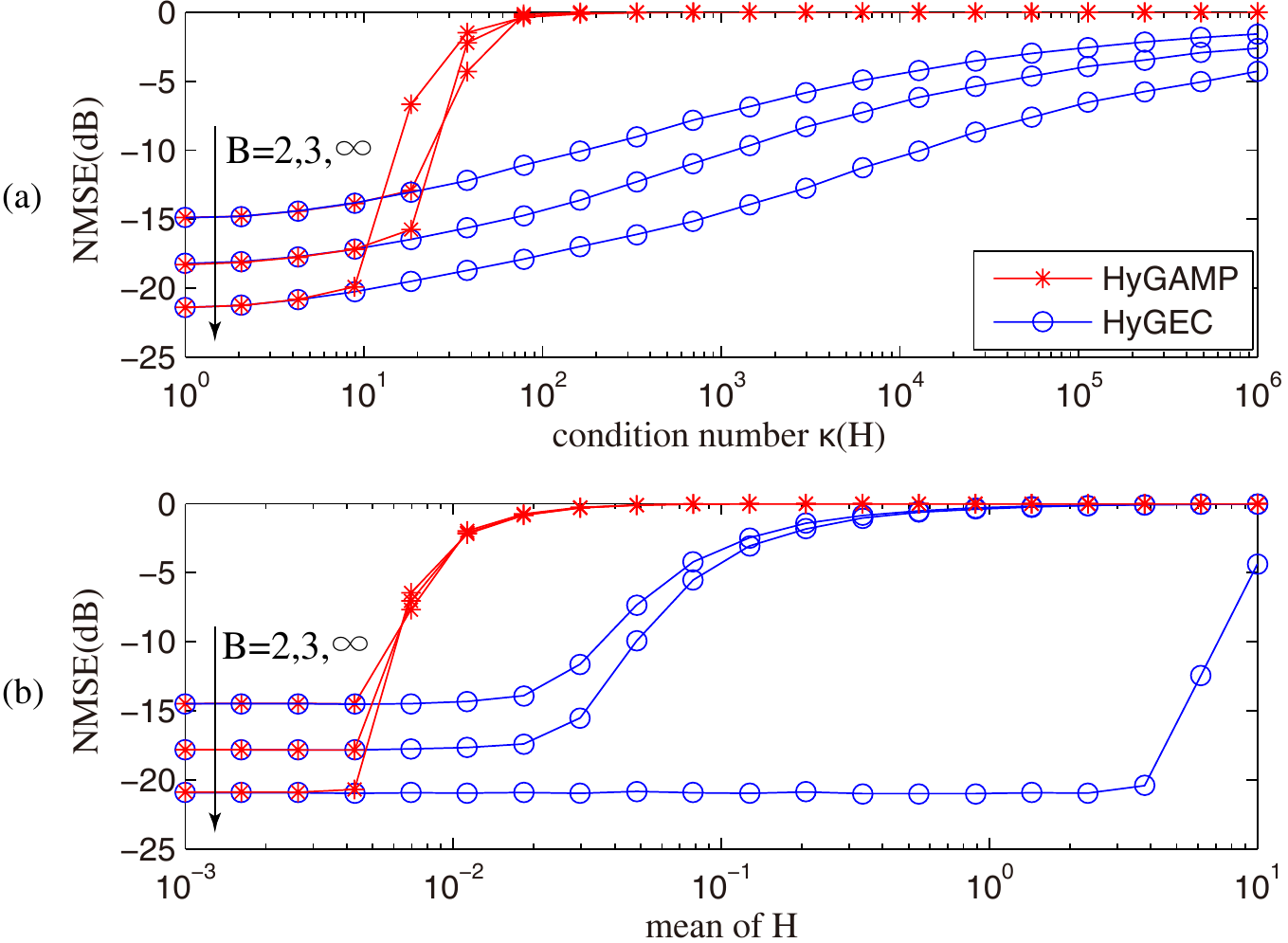}
\caption{(a) condition number $\kappa(\bs{H})$ versus NMSE of $\bs{x}$; (b) the mean of $\bs{H}$ versus NMSE of $\bs{x}$. The B refers to the number of ADC bits.
}
\label{Fig:Sim2}
\end{figure}

In Fig.~\ref{Fig:Sim1}, we give the iteration versus normalized MSE (NMSE) of $\bs{x}$ ($\mathbb{E}\{\|\hat{\bs{x}}-\bs{x}\|^2\}/\mathbb{E}\{\|\bs{x}\|^2\}$) of HyGEC algorithm (known $\rho$) and EM-aided HyGEC (EM to learn $\rho$). The dimensions of system are set as $(M,N,K,\rho)=(1000,2000, 100, 0.1)$ and the signal-to-noise rate (SNR) is set as $10$dB. In EM-aided HyGEC, the sparse rate is initialized as $\rho=0.01$ and then outer iteration EM learns the actual sparse rate. As can be seen from Fig.\ref{Fig:Sim1}, the EM-aided HyGEC algorithm can attain the same performance as HyGEC algorithm when they converge.

In Fig.~\ref{Fig:Sim2}, the dimensions of system are set as $(M,N,K,\rho)$ $=(500, 1000, 100, 0.1)$ and the SNR is $12$dB. In Fig.~\ref{Fig:Sim2}(a), we present the plot of the condition number $\kappa(\bs{H})$ versus NMSE of $\bs{x}$, where  $\kappa(\bs{H})=\frac{\max(\text{eig}(\bs{H}))}{\min(\text{eig}(\bs{H}))}$.  Let $\bs{H}=\bs{U}\boldsymbol{\Sigma}\bs{V}^{\text{T}}$, where $\bs{U}$ and $\bs{V}$ are independent Haar-distributed matrices. The non-zero singular values are set as $\lambda_{i+1}/\lambda_i=\varrho$, $\varrho^{M}=\kappa(\bs{H})$, and $\sum_{i=1}^M\lambda_i^2=M$. From this sub-figure, it can be found that the NMSE performance of the proposed HyGEC algorithm is equal to that of HyGAMP in low $\kappa(\bs{H})$. In addition, the HyGAMP algorithm fails to converge in $\kappa(\bs{H})>10^2$, but HyGEC can also converge with its NMSE performance suffering a degradation. In Fig.~\ref{Fig:Sim2} (b), we give the plot of  the mean of $\bs{H}$ versus NMSE of $\bs{x}$. As can be seen from sub-figure (b), the HyGEC is relatively insensitive to HyGAMP for $\bs{H}$ with non-zero mean.

\section{Conclusion}
This paper considered the signal recovery of structured sparse signal of generalized linear model in which the activity of each group is decided by a hidden binary variable. To solve this problem, we proposed a novel algorithm called hybrid generalized expectation consistent (HyGEC) based on the hybrid factor graph. The hybrid factor graph can be divided into two parts: traditional factor graph (scalar) and vector factor graph. The sum-product loopy belief propagation (LBP) is run in traditional factor graph part while the EP manner is implemented in the vector factor graph part. Additionally, to learn the unknown sparse rate we apply EM algorithm together with the HyGEC algorithm where the HyGEC is run as the E-step of EM algorithm. Finally, the simulation results verify that the proposed algorithm holds on more general regions than the competing HyGAMP algorithm.

\begin{appendices}
\section{Derivation of HyGEC}
\label{Sec1}

\begin{table}[h]
\centering
\caption{Notation definitions for messages}
\label{Table_Definition}
\begin{tabular}{|l|l|l|l|l|}	
\hline
$\mu_{k\leftarrow j}^{\xi}(\xi_k,t)$    &message from $\mathcal{P}(x_{kj}|\xi_k)$ to  $\xi_k$ \\
\hline
$\mu_{k\rightarrow j}^{\xi}(\xi_k,t)$   &message from $\xi_k$ to $\mathcal{P}(x_{kj}|\xi_k)$\\
\hline
$\mu^{\text{pri}}(\bs{z},t)$             &message from $\bs{z}$ to $\mathcal{P}(\bs{y}|\bs{z})$\\
\hline
$\mu^{\text{lik}}(\bs{z},t)$              &message from $\mathcal{P}(\bs{y}|\bs{z})$ to $\bs{z}$ \\
\hline
$\mu^{\text{lik}}(\bs{x},t)$              &message from $\bs{x}$ to $\mathcal{P}(\bs{x}|\boldsymbol{\xi})$  \\
\hline
$\mu^{\text{pri}}(\bs{x},t)$             &message from $\mathcal{P}(\bs{x}|\boldsymbol{\xi})$ to $\bs{x}$ \\
\hline
$\mu^{\xi}(\xi_k,t)$                     &belief distribution of $\xi_k$\\
\hline
\end{tabular}
\end{table}

\begin{figure}
	\centering
	\includegraphics[width=0.65\textwidth]{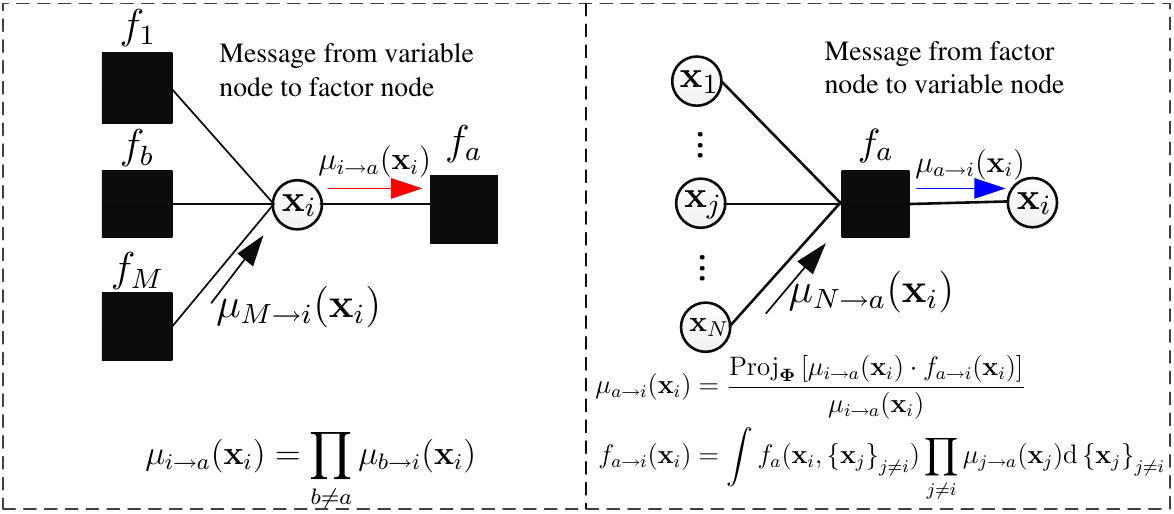}
	\caption{The message update rules of EP manner \cite{zou2018concise}, \cite{meng2015expectation}.
	}
	\label{Fig2:Rules}
\end{figure}

The derivation of messages of part $\mathcal{P}(\boldsymbol{\xi})\leftrightarrow\boldsymbol{\xi}\leftrightarrow\mathcal{P}(\bs{x}|\boldsymbol{\xi})$ is the same as the LBP part in \cite{zou2020message}. We here only give the derivation of messages in vector factor graph part. Using EP update rules in Fig.~\ref{Fig2:Rules}, we address the following messages defined in Table~\ref{Table_Definition}. Note that we omit the iteration time.
\begin{subequations}
	\begin{align}
	\!\!\mu^{\text{lik}}(\bs{z})&\propto \frac{\text{Proj}_{\bs{\Phi}}[ \mathcal{P}(\bs{y}|\bs{z})\mu^{\text{pri}}(\bs{z})]}{\mu^{\text{pri}}(\bs{z})},
	\label{Equ:V1}\\
	\!\!\mu^{\text{lik}}(\bs{x})&\propto  \frac{\text{Proj}_{\boldsymbol{\Phi}}\left[\int \mu^{\text{pri}}(\bs{x})\delta(\bs{z}-\bs{Hx})\mu^{\text{lik}}(\bs{z})\text{d}\bs{z}\right] }{\mu^{\text{pri}}(\bs{x})},
	\label{Equ:V2}\\
	\!\!\mu^{\text{pri}}(\bs{x})&\propto \frac{\text{Proj}_{\boldsymbol{\Phi}}\left[\prod\limits_{k=1}^K\prod\limits_{j=1}^{N_k}\int \mu_{k\rightarrow j}^{\xi}(\xi_k) \mathcal{P}(x_{kj}|\xi_{k})\text{d}\xi_k\mu^{\text{lik}}(\bs{x})\right]}{\mu^{\text{lik}}(\bs{x})},
	\label{Equ:V3}\\
	\!\!\mu^{\text{pri}}(\bs{z})&\propto \frac{\text{Proj}_{\boldsymbol{\Phi}}\left[\int \!\!\mu^{\text{pri}}(\bs{x},t)\delta(\bs{z}-\bs{Hx})\mu^{\text{lik}}(\bs{z})\text{d}\bs{x}\right]}{\mu^{\text{lik}}(\bs{z})},
	\label{Equ:V4}
	\end{align}
\end{subequations}
where the superscript `$\text{pri}$' means approximate prior while `$\text{lik}$' denotes approximate likelihood function, and
\begin{align}
\text{Proj}_{\bs{\Phi}}[q(\bs{x})]=\mathcal{N}(\bs{x}|\bs{m},\bs{v})
\end{align}
with $\bs{m}=\int \bs{x}q(\bs{x})\text{d}\bs{x}$ and $\bs{v}=\int \|\bs{x}-\bs{m}\|^2q(\bs{x})\text{d}\bs{x}$.

It is assumed that each message in EP manner is Gaussian distribution. We first calculate the term in (\ref{Equ:V1})
\begin{align}
\mathcal{N}(\bs{z}|\hat{\bs{z}}^{\text{pos}}, \bs{v}^{\text{pos}}_{\text{z}})&=\text{Proj}_{\bs{\Phi}}[ \mathcal{P}(\bs{y}|\bs{z})\mu^{\text{pri}}(\bs{z})]
\end{align}
where $\hat{\bs{z}}^{\text{pos}}=\mathbb{E}\{\bs{z}|\bs{m}_{\text{z}}^{\text{pri}},\bs{v}_{\text{z}}^{\text{pri}}\}, \bs{v}_{\text{z}}^{\text{pos}}=\text{Var}\{\bs{z}|\bs{m}_{\text{z}}^{\text{pri}},\bs{v}_{\text{z}}^{\text{pri}}\}$ with $\mu^{\text{pri}}(\bs{z})=\mathcal{N}(\bs{x}|\bs{m}_{\text{z}}^{\text{pri}},\bs{v}_{\text{z}}^{\text{pri}})$ and the expectation over $\frac{\mathcal{P}(\bs{y}|\bs{z})\mathcal{N}(\bs{z}|\bs{m}_{\text{z}}^{\text{pri}},\bs{v}_{\text{z}}^{\text{pri}})}{\int \mathcal{P}(\bs{y}|\bs{z})\mathcal{N}(\bs{z}|\bs{m}_{\text{z}}^{\text{pri}},\bs{v}_{\text{z}}^{\text{pri}})\text{d}\bs{z}}$.

By Gaussian reproduction property, from (\ref{Equ:V1}) we have $\mu^{\text{lik}}(\bs{z})=\mathcal{N}(\bs{z}|\bs{m}_{\text{z}}^{\text{lik}},\bs{v}_{\text{z}}^{\text{lik}})$ where
$
\bs{v}_{\text{z}}^{\text{lik}}=\bs{1}\oslash (\bs{1}\oslash \bs{v}_{\text{z}}^{\text{pos}}-\bs{1}\oslash \bs{v}_{\text{z}}^{\text{pri}}),\
\bs{m}_{\text{z}}^{\text{lik}}=\bs{v}_{\text{z}}^{\text{lik}}\odot (\hat{\bs{z}}\oslash \bs{v}_{\text{z}}^{\text{pos}}-\bs{m}_{\text{z}}^{\text{pri}}\oslash \bs{v}_{\text{z}}^{\text{pri}})$,
where $\oslash$ and $\odot$ denote the element-wise multiplication and division, respectively.

We move to the computation of (\ref{Equ:V2}). The projection term in (\ref{Equ:V2}) is
\begin{align}
\!\!\!\!\!\!\mathcal{N}(\bs{x}|\hat{\bs{x}}^{\text{pos2}},\bs{v}_{\text{x}}^{\text{pos2}})
&=\text{Proj}_{\boldsymbol{\Phi}}[\int \mu^{\text{pri}}(\bs{x})\delta(\bs{z}-\bs{Hx})\mu^{\text{lik}}(\bs{z})\text{d}\bs{z}].
\end{align}
By Gaussian reproduction property, we obtain
\begin{align}
\bs{Q}_{\text{x}}&=\left(\bs{H}^{\text{T}}\text{Diag}(\bs{1}\oslash \bs{v}^{\text{lik}}_{\text{z}})\bs{H}+\text{Diag}(\bs{1}\oslash \bs{v}_{\text{x}}^{\text{pri}})\right)^{-1},\\
\hat{\bs{x}}^{\text{pos2}}&=\bs{Q}_{\text{x}}\left(\bs{H}^{\text{T}}\text{Diag}(\bs{1}\oslash \bs{v}^{\text{lik}}_{\text{z}})\bs{m}^{\text{lik}}_{\text{z}}+\bs{m}_{\text{x}}^{\text{pri}}\oslash \bs{v}_{\text{x}}^{\text{pri}}\right),\\
\bs{v}_{\text{x}}^{\text{pos2}}&=\text{diag}(\bs{Q}_{\text{x}}),
\end{align}
where $\text{diag}(\bs{A})$ denotes a vector whose elements are from the diagonal elements of square matrix $\bs{A}$, while $\text{Diag}(\bs{a})$ denotes a square matrix whose diagonal element is $\bs{a}$.

From (\ref{Equ:V2}), we have $\mu^{\text{lik}}(\bs{x})=\mathcal{N}(\bs{x}|\bs{m}_{\text{x}}^{\text{lik}},\bs{v}_{\text{x}}^{\text{lik}})$ whose mean and variance are given by $
\bs{v}_{\text{x}}^{\text{lik}}=\bs{1}\oslash (\bs{1}\oslash \bs{v}_{\text{x}}^{\text{pos2}}-\bs{1}\oslash \bs{v}_{\text{x}}^{\text{pri}}),
\bs{m}_{\text{x}}^{\text{lik}}=\bs{v}_{\text{x}}^{\text{lik}}\odot (\hat{\bs{x}}^{\text{pos2}}\oslash \bs{v}_{\text{x}}^{\text{pos2}}-\bs{m}_{\text{x}}^{\text{pri}}\oslash \bs{v}_{\text{x}}^{\text{pri}})
$.

For (\ref{Equ:V3}), we first calculate the following term. For easy of notation, we define
\begin{align}
\!\!\!\!\! \hat{\mathcal{P}}(x_{kj})
&=\int \mu_{k\rightarrow j}^{\xi}(\xi_k,t) \mathcal{P}(x_{kj}|\xi_{k})\text{d}\xi_k\\
&=\mu_{k\rightarrow j}^{\xi}(\xi_k=1)\mathcal{P}_{\textsf{X}}(x_{kj})+\mu_{k\rightarrow j}^{\xi}(\xi_k=0)\delta(x_{kj})\\
&=\hat{\rho}_{kj}\mathcal{P}_{\textsf{X}}(x_{kj})+(1-\hat{\rho}_{kj})\delta(x_{kj}),
\label{Equ:xpri}
\end{align}
where the definition $\hat{\rho}_{kj}=\mu_{k\rightarrow j}^{\xi}(\xi_k=1)$ is applied, which can also be written as
\begin{align}
\hat{\rho}_{kj}&=1-\frac{1}{1+\exp (\text{LLR}_{k\rightarrow j}^{\xi})}.
\end{align}

We denote
\begin{align}
\mathcal{N}(\bs{x}|\hat{\bs{x}}^{\text{pos}},\bs{v}_{\text{x}}^{\text{pos}})&=\text{Proj}_{\boldsymbol{\Phi}}\left[\hat{\mathcal{P}}(\bs{x})\mu^{\text{lik}}(\bs{x})\right]
\end{align}
where $\hat{\bs{x}}^{\text{pos}}=\mathbb{E}\left\{\bs{x}|\bs{m}_{\text{x}}^{\text{lik}},\bs{v}_{\text{x}}^{\text{lik}};\hat{\boldsymbol{\rho}}\right\}
, \bs{v}_{\text{x}}^{\text{pos}}=\text{Var}\left\{\bs{x}|\bs{m}_{\text{x}}^{\text{lik}},\bs{v}_{\text{x}}^{\text{lik}};\hat{\boldsymbol{\rho}}\right\}$, where the expectation is taken over $\frac{\hat{\mathcal{P}}(\bs{x})\mathcal{N}(\bs{x}|\bs{m}_{\text{x}}^{\text{lik}},\bs{v}_{\text{x}}^{\text{lik}})}{\int \hat{\mathcal{P}}(\bs{x})\mathcal{N}(\bs{x}|\bs{m}_{\text{x}}^{\text{lik}},\bs{v}_{\text{x}}^{\text{lik}}) \text{d}\bs{x}}$ with $\mu^{\text{lik}}(\bs{x})=\mathcal{N}(\bs{x}|\bs{m}_{\text{x}}^{\text{lik}},\bs{v}_{\text{x}}^{\text{lik}})$ and $\hat{\mathcal{P}}(\bs{x})$ is found in (\ref{Equ:xpri}).

By (\ref{Equ:V3}), we obtain $\mu^{\text{pri}}(\bs{x})$ as $\mathcal{N}(\bs{x}|\bs{m}_{\text{x}}^{\text{pri}},\bs{v}_{\text{x}}^{\text{pri}})$, where the mean and variance are given by
$
\bs{v}_{\text{x}}^{\text{pri}}=\bs{1}\oslash (\bs{1}\oslash \bs{v}_{\text{x}}^{\text{pos}}-\bs{1}\oslash  \bs{v}_{\text{x}}^{\text{lik}}),\
\bs{m}_{\text{x}}^{\text{pri}}=\bs{v}_{\text{x}}^{\text{pri}}\odot (\hat{\bs{x}}^{\text{pos}}\oslash \bs{v}_{\text{x}}^{\text{pos}}-\bs{m}_{\text{x}}^{\text{lik}}\oslash  \bs{v}_{\text{x}}^{\text{lik}})
$.

For (\ref{Equ:V4}), we calculate
\begin{align}
\int \mu^{\text{pri}}(\bs{x},t)\delta(\bs{z}-\bs{Hx})\mu^{\text{lik}}(\bs{z},t)\text{d}\bs{x}
&\propto \int \delta(\bs{z}-\bs{Hx})\mathcal{N}(\bs{x}|\hat{\bs{x}}^{\text{pos}2},\bs{Q}_{\text{x}})\text{d}\bs{x}\\
&=\mathcal{N}(\bs{z}|\hat{\bs{z}}^{\text{pos1}},\bs{v}^{\text{pos1}}_{\text{z}}),
\end{align}
where the last equation holds by PDF-to-RV lemma\footnote{
	Let ${\bs{w}}\in \mathbb{R}^a$ and $\bs{u} \in \mathbb{R}^b$ be two RVs, and $g:\mathbb{R}^a\rightarrow \mathbb{R}^b$ be a generic mapping. Then, $\bs{u}=g({\bs{w}})$ if and only if the PDF $\mathcal{P}(\bs{u})\propto \int \delta(\bs{u}-g(\bs{w}))\mathcal{P}(\bs{w})\text{d}\bs{w}$.}, and
\begin{align}
\hat{\bs{z}}^{\text{pos1}}&=\bs{H}\hat{\bs{x}}^{\text{pos}2},\\
\bs{v}^{\text{pos1}}_{\text{z}}&=\text{diag}(\bs{H}\bs{Q}_{\text{x}}\bs{H}^{\text{T}}).
\end{align}

Similarly, from (\ref{Equ:V4}), we obtain
$
\bs{v}_{\text{z}}^{\text{pri}}=\bs{1}\oslash (\bs{1}\oslash \bs{v}^{\text{pos1}}_{\text{z}}-\bs{1}\oslash \bs{v}_{\text{z}}^{\text{lik}}),\
\bs{m}_{\text{z}}^{\text{pri}}=\bs{v}_{\text{z}}^{\text{pri}}\odot (\hat{\bs{z}}^{\text{pos1}}\oslash \bs{v}^{\text{pos1}}_{\text{z}}-\bs{m}_{\text{z}}^{\text{lik}}\oslash \bs{v}_{\text{z}}^{\text{lik}}).
$

\end{appendices}

\bibliographystyle{IEEEtran}%
\bibliography{ZQY_bib}

\
\end{document}